\documentclass[3p,times,twocolumn]{elsarticle}

\usepackage{ecrc}
\usepackage{graphicx}
\usepackage{subfig}

\volume{00}

\firstpage{1}

\journalname{Nuclear Physics B Proceedings Supplement}

\runauth{Donglian Xu}

\jid{nuphbp}

\jnltitlelogo{Nuclear Physics B Proceedings Supplement}

\usepackage{amssymb}
\usepackage{hyperref}

 \usepackage{lineno}

\usepackage[figuresright]{rotating}

\begin{document}

\begin{frontmatter}



\dochead{}

\title{Exploring the Universe with Neutrinos:\\ Recent Results from IceCube}


%\author{Donglian Xu\footnotemark[0] for the IceCube Collaboration\footnotemark[1]}
\author{Donglian Xu\corref{cor1} for the IceCube Collaboration\fnref{label1}}

\cortext[cor1]{Email: dxu@icecube.wisc.edu}

\fntext[label1]{http://icecube.wisc.edu/}

\address{Wisconsin Particle Astrophysics Center, 222 West Washington Ave., Suite 500
Madison, WI 53703, USA \\  Department of
  Physics, University of Wisconsin-Madison, 1150 University Avenue
Madison, WI  53706, USA
}
%\email{donglian.xu@icecube.wisc.edu}

%\footnotemark[0]{donglian.xu@icecube.wisc.edu}

\begin{abstract}
In 2013, the IceCube Neutrino Observatory located at the geographic South Pole
detected evidence for a diffuse astrophysical neutrino flux above
$\sim$60 TeV.
%%, which is inconsistent with atmospheric origin. 
To this day, IceCube has
operated with full detector configuration for more than 6 years. 
The observed astrophysical neutrino flux has been confirmed with $> 6
\sigma$ significance with both events starting within the
detector (all flavor) and events traversing through the Earth ($\nu_{\mu}$
charged-current). 
Somewhat equal flavor ratio of astrophysical neutrinos is expected at
Earth assuming standard thorough oscillation. 
%among the neutrinos that have traversed astronomical distances. 
%An equal flavor ratio of astrophysical neutrinos is expected at Earth in the standard pion-production and neutrino oscillation scenarios.
%Identification of tau neutrinos in this flux have been carried out but
%yielded null result.
A search for tau neutrinos has been carried out but yielded null result. 
No neutrino sources have been found to contribute
significantly to the diffuse flux at this point. In this paper, we
will review the current status of the astrophysical neutrino flux, discuss the quest for neutrino point
sources and overview the proposed design and physics potentials of the future IceCube-Gen2. 
\end{abstract}


\end{frontmatter}


%%\linenumbers

\section{Cosmic Rays, Cosmic Neutrinos and IceCube}
\label{s1}

Charged particles from space known as cosmic rays are constantly bombarding the
Earth's atmosphere, fragmenting air nuclei and producing copious
hadrons which subsequently decay into $\gamma$-rays, muons and neutrinos. 
The energy spectrum of the cosmic rays detected at
Earth spans greater than 10 orders of magnitude, and follows an almost
featureless power law of $E^{-2.7}$ from $\sim$10$^9$ eV to beyond $10^{20}$
eV. Even after a century since their discovery, the origin of cosmic
rays are largely still unknown.
%The origin of cosmic rays is still largely unknown even after a
%century-long discovery. 
%It is expected that in the
%intermediate energys region ($<10^{16}$ eV), the
%dominant contribution of cosmic rays come from the local Galaxy, while
%the ultra-high-energy cosmic rays (UHECRs, $>10^{18}$ eV) are predominantly originated from extra-galactic
%sources. Observations show that the power law energy spectrum of cosmic rays has
%a steepening at $\sim 3 \times 10^{15} $ eV, indicating an acceleration
%power limit from the Galaxy. Highly magnetized supernova remnants are
%confirmed to be accelerators of relatively low energetic cosmic rays
%in the GeV energy range, from the detection of $\pi^{\circ}
%\rightarrow 2\gamma$ characteristics which is an indicator of hadronic
%processes \cite{Ackermann807}. 
It is expected that the ultra-high-energy
($>10^{18}$ eV) cosmic rays (UHECRs) are accelerated by extremely energetic engines and events such as active galactic
nuclei (AGN) and gamma ray bursts (GRBs). Within the acceleration
sites, $\gamma$-rays and neutrinos are produced from protons interacting
with the ambient materials via the $pp$ and $p\gamma$ hadronic processes. 
Not only are UHECRs deflected by magnetic fields, which make it
difficult to relate back to their origin, but they also can interact with
the cosmic microwave background (CMB) via the $p\gamma \rightarrow \Delta^{+}$ process,
confining them to a horizon $\sim$50~Mpc.
%\cite{GZK}. 
High energy ($>$PeV) photons, produced in distant cosmic ray acceleration sites
interact with the extragalactic background light, which also limits
the visibility to those potential accelerators in the $\gamma$ channel.
Neutrinos, on the other hand, have no charge and interact only
weakly. They can traverse astronomical distances without being
absorbed and point back to
their origin once detected on Earth. They are a direct diagnostics of hadronic processes.
Therefore, detection of high energy neutrino sources will be smoking gun evidence for
UHECRs accelerators. 
%The neutrino flux from AGNs is estimated to be
%E$^2\Phi$ $\sim 10^{-8}$ GeV cm$^{2}$ sr$^{-1}$ s$^{-1}$ assuming the
%sources are transparent to photons, known as the Waxman-Bahcall upper
%bound \cite{Waxman-Bahcall}. A large detector is required to detect
%such a low flux of astrophysical neutrinos with small cross sections.   

The IceCube Neutrino Observatory is a cubic-kilometer neutrino
detector located at the geographic South Pole. It was built to be
sensitive to astrophysical neutrinos in the TeV to PeV energies.
The construction of IceCube started
in 2004 and completed in December 2010. IceCube consists of an array of 5160 digital optical modules
(DOMs) deployed in the Antarctic glacial ice cap at depths between
1450~m and 2450~m from the surface, a straight overburden of 1.5~km of
ice to shield the down-going atmospheric muons. There are 86 cables
called {\it strings}, each of which has 60 DOMs. The inter-string
distance is $\sim$125~m, and the vertical distance between
two DOMs is $\sim$17~m. A more densely (inter-string 60-70~m,
inter-DOM 7~m) instrumented sub-array called
DeepCore is located at the bottom center of IceCube, which lowers the detection
energy threshold to $\sim$10~GeV and opens a window for atmospheric
neutrino oscillation physics and new physics at these energies. 
%The inter string distance for DeepcCore is $\sim$ 60-70m, and the
%vertical distance between DOMs is $\sim$ 7m. 

Neutrinos cannot be detected
directly. In IceCube, neutrinos are detected via
the Cherenkov photons emitted by the relativistic secondary particles from neutrino
interacting with the ice nuclei. It relies on the precise
reconstruction of such interactions based on the timing and
charge information collected by each DOM during an event
readout. There are two main categories of event topologies in IceCube:
{\it tracks} and {\it cascades}. A track is made by a
($\nu_{\mu}$-induced) muon traveling through the detector and emitting
Cherenkov radiation 
%plus stochastic energy losses 
along its trajectory. A cascade can be made by neutral current (NC) interactions of
all neutrino flavors, $\nu_e$ charged current (CC) interactions and
low energy $\nu_{\tau}$ CC interactions. During these interactions, relativistic
hadrons and electrons are created, then decay or interact,
producing a shower of subsequent particles. Angular resolution for
tracks is $<1^{\circ}$ and for cascades is $\sim$15$^{\circ}$ above
100~TeV. Energy resolution for tracks is about a factor of 2, while for
cascades is $\sim$10\% \cite{Aartsen:2013vja}. 
A third type of event
topology called {\it double cascade} (not yet observed) can be made by high energy
$\nu_{\tau}$ undergoing CC interaction, producing one cascade, and the
outgoing $\tau$ lepton decays subsequently into hadrons or electrons,
producing a second cascade.   

 \section{Diffuse Astrophysical Neutrino Flux}
 \label{s2}

\begin{figure}[h]
\includegraphics[width=0.45\textwidth]{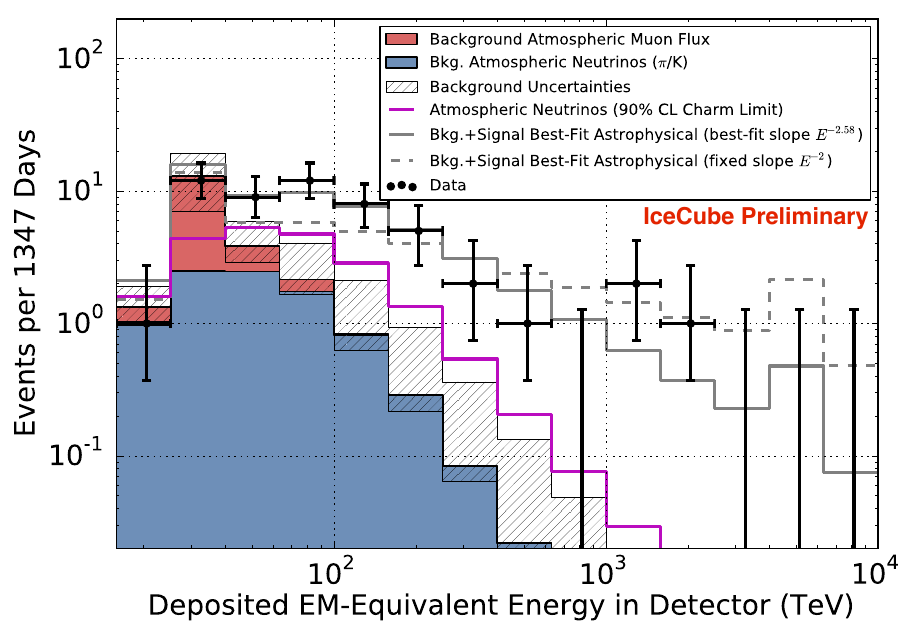}
\caption{Deposited energies distribution of the 54 starting events 
  selected by a veto method in 4 years of data. Black crosses
  represent data points, the
  grey solid line is the best fit of astrophysical neutrino flux with
  an unbroken power law \cite{Kopper:2015vzf}. }
\label{fig:hese_4yr}
\end{figure}

The IceCube detector is triggered by down-going atmospheric muons at a
rate of $\sim$3~kHz. There is roughly one atmospheric neutrino in a million
cosmic-ray induced muons, while this number for astrophysical
neutrinos is one in a billion. It is, therefore, challenging to separate the astrophysical neutrinos
from the enormous atmospheric backgrounds. Two PeV cascade events were first discovered
in an analysis optimized for extremely high energy (EHE) GZK neutrino search, hinting
astrophysical neutrinos at 2.8$\sigma$ significance above
atmospheric backgrounds \cite{Aartsen:2013bka}.
Currently, two strategies are employed to sufficiently select astrophysical
neutrinos. One is using a thin outer layer of the detector as an active veto to
reject incoming muons, and keeping only events starting within
the detector fiducial volume. This method is very efficient to
select high energy neutrino events as most high energy muons will leave light along
their trajectory when passing through the detector, while
neutrinos only produce light when their interactions with ice nuclei
occur. This veto method is sensitive to neutrinos of all flavors from
the whole sky. With an event-wise charge cut of 6000 photoelectrons, 
an active veto method was rapidly developed to find more high energy
starting events (HESE) after the two PeV cascades were found. 
With the veto method, a diffuse astrophysical neutrino
flux has been observed at $>$ 6$\sigma$ significance above $\sim$60~TeV
\cite{Kopper:2015vzf, Aartsen:2013jdh, Aartsen:2014gkd}. 
A total of 54 events were detected in 1347 days from
May 2010 to May 2014 (Fig.~\ref{fig:hese_4yr}), with expected
$12.6 \pm 5.1$ atmospheric muons and $9.0^{+8.0}_{-2.2}$ atmospheric
neutrinos. The best fit all-flavor flux assuming an unbroken power law
is $\Phi_{\nu+\bar\nu} = 2.2 \pm 0.7 \cdot(\frac{E_{\nu}}{100~\mathrm{TeV}})^{-(2.58 \pm 0.25)} \times10^{-18}$ GeV$^{-1}$ cm$^{-2}$ s$^{-1}$
sr$^{-1}$ \cite{Kopper:2015vzf}. 
An adaptive veto which expands as
energy lowers (more veto volume is needed to tag dimmer muons) was
developed to find even more starting events above $\sim$1~TeV, resulted in
consistent results \cite{Aartsen:2014muf}.
\begin{figure*} [t]
\centering
 %%\subfloat{{\includegraphics[width=0.47\textwidth]{IC59_align_new_IC79_align_new_IC86-2011_align_new_IC86-2012-13-14-15_wChargeCorrection_align_bestfit_fluxes_timesE2_wHESE.png}}}%
 \subfloat{{\includegraphics[width=0.47\textwidth]{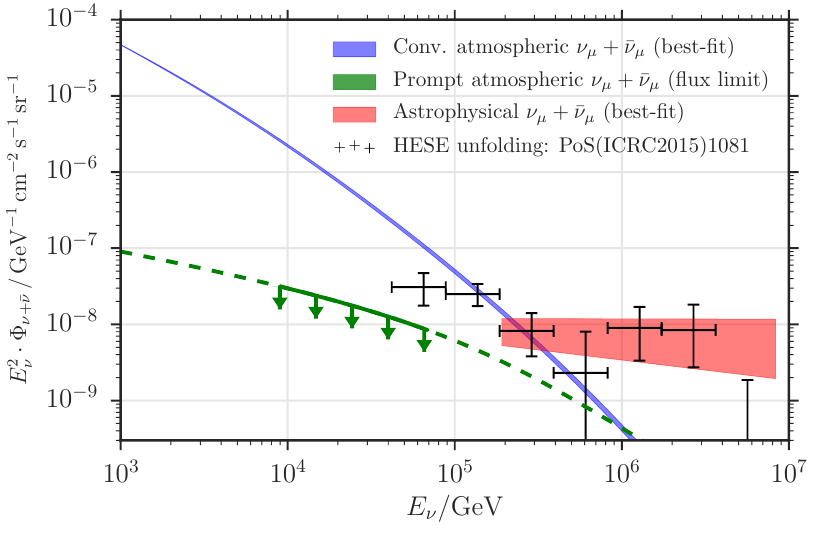} }}%
    \qquad
    %\subfloat{{\includegraphics[width=0.47\textwidth]{Skymap_with_energy_milli_IC2015.png} }}%
    \subfloat{{\includegraphics[width=0.47\textwidth]{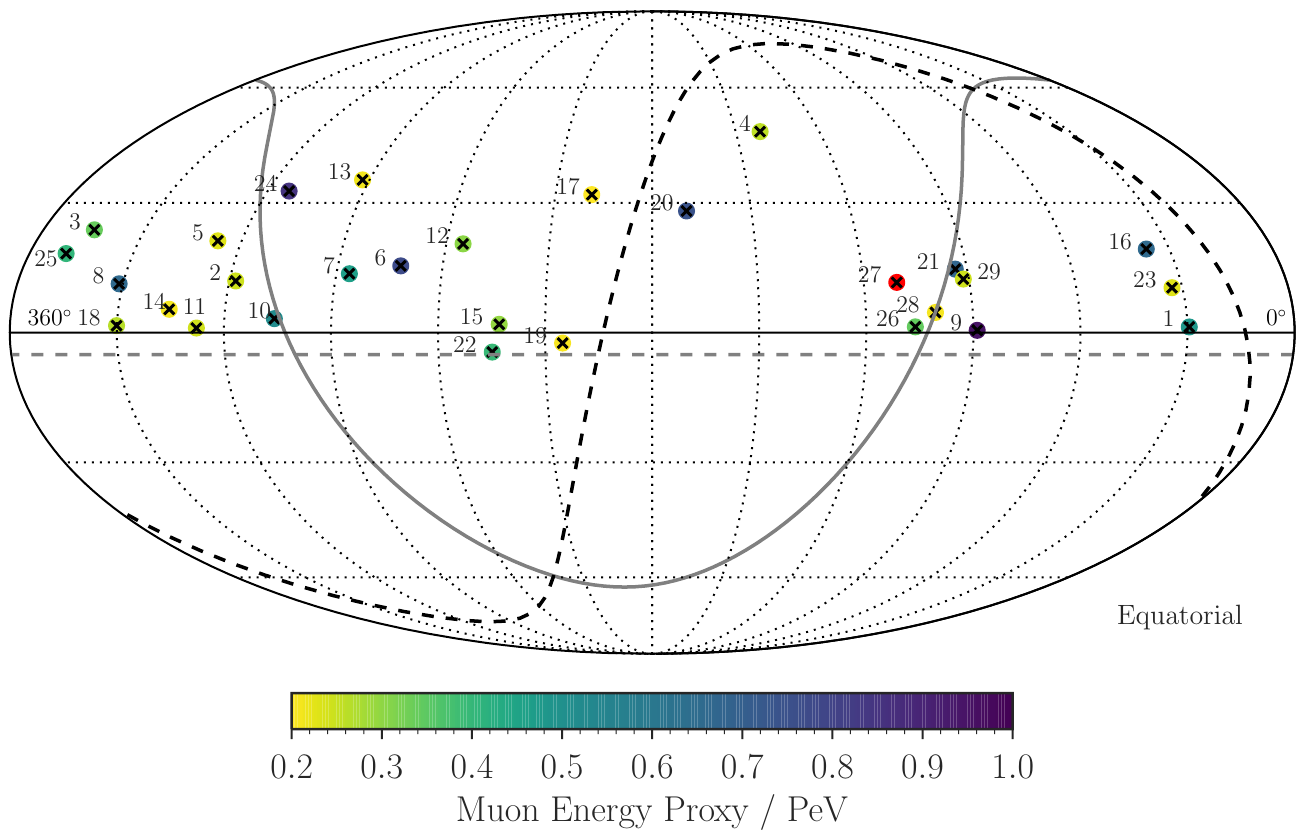} }}%
\caption{Left: best fit of astrophysical $\nu_{\mu}+\bar\nu_{\mu}$
  flux (red band). Right: sky map of 29 events with reconstructed
  muon energy proxy $>200$ TeV from 6 years of through-going track sample;
  the signal to noise ratio is $\sim$ 2:1 \cite{Aartsen:2016xlq}.}%
\label{fig:numu_6yr}%
\end{figure*}
%Another strategy is to use the Earth as a filter for atmospheric muons,
%and obtain only up-going track-like events, which are
%predominantly $\nu_{\mu}$ CC events and a small fraction of
%$\nu_{\tau}$ CC events with the $\tau$ lepton decaying into muons. 
Another strategy is to use the Earth as a filter for atmospheric
muons, leaving only up-going track-like events.  
These are predominantly $\nu_{\mu}$ CC events and a small fraction of
$\nu_{\tau}$ CC events with the $\tau$ lepton decaying into muons.
This search measured an astrophysical $\nu_{\mu}$ flux at 3.8$\sigma$ in two
years of IceCube data from 2010 to 2012 \cite{Aartsen:2015rwa}. To date, 6
years of data has been analyzed and the significance has reached
5.6$\sigma$. The unbroken power law best fit of the astrophysical
$\nu_{\mu} + \bar\nu_{\mu}$ flux is $\Phi_{\nu+\bar\nu} = 0.90^{+0.30}_{-0.27} \cdot
(\frac{E_{\nu}}{100~\mathrm{TeV}})^{-(2.13 \pm 0.13)} \times 10^{-18}$
GeV$^{-1}$ cm$^{-2}$ sr$^{-1}$ s$^{-1}$, consistent with the all-sky all
flavor flux measured from starting events, as shown in left panel of
Fig.~\ref{fig:numu_6yr}. Among the total 352,294 events, 
there are 29 events with
reconstructed muon energy proxy greater than 200 TeV, with the highest at $2.6\pm
0.3$ PeV. Their arrival directions are projected onto the sky as shown in the right panel of
Fig.~\ref{fig:numu_6yr}. No spatial or timing clustering was found among
these events, neither did they correlate with any gamma ray source
catalogs considered \cite{Aartsen:2016xlq}.

\begin{figure}[th]
\includegraphics[width=0.45\textwidth]{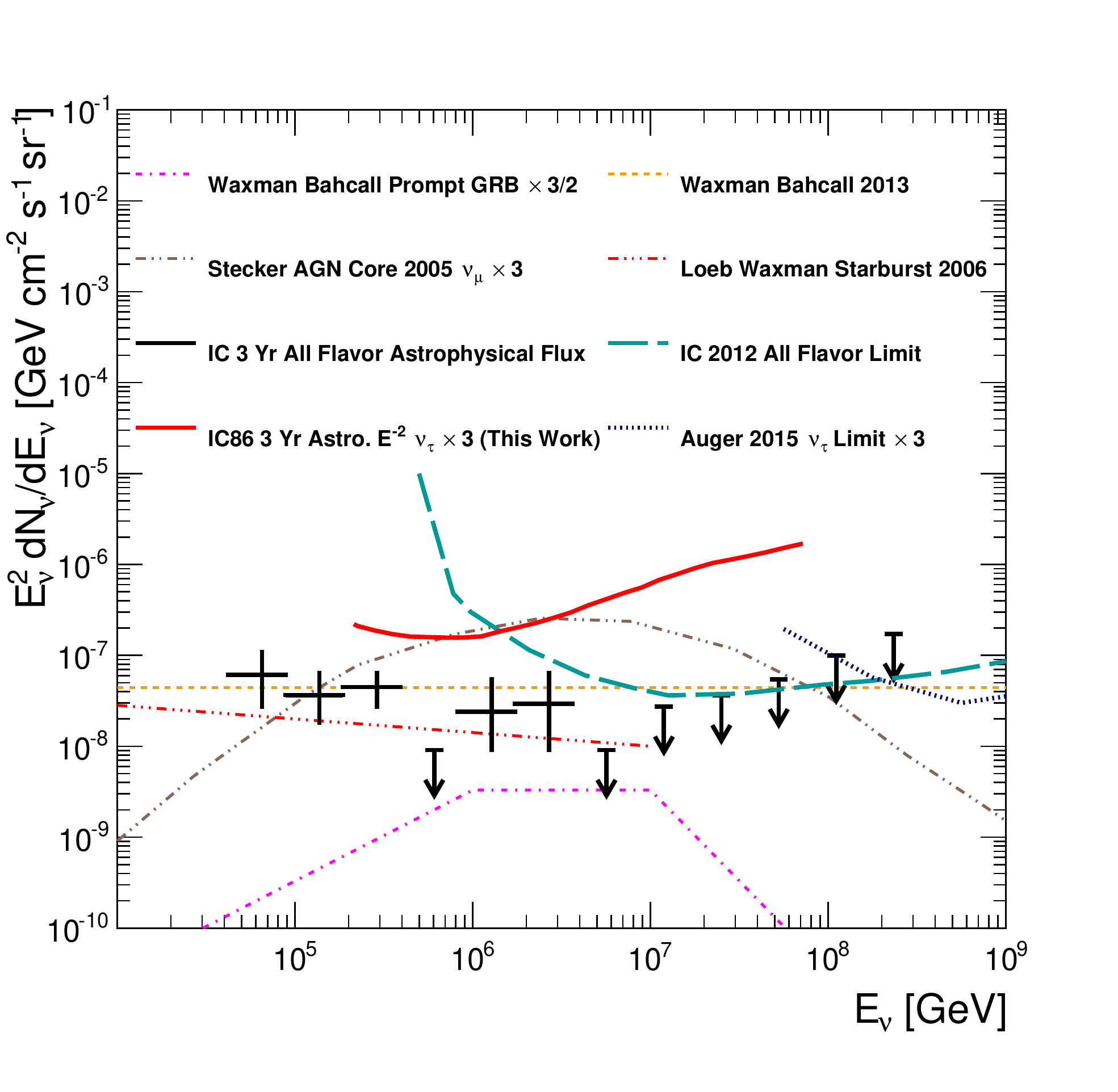}
\caption{Differential upper limits (red) on astrophysical tau neutrino flux
  based on null results from a search using three years of data \cite{Aartsen:2015dlt}.}
\label{fig:3yr_tau}
\end{figure}

Astrophysical neutrinos at Earth are expected to have somewhat equal flavor
ratios due to standard thorough oscillation. Therefore, there should be
a large number of tau neutrinos in the observed flux. At energies
below PeV, the double cascade topology from $\nu_{\tau}$ CC
interactions is very difficult to distinguish from single cascades,
given that the separation distance scales as ($\frac{E_{\tau}}{1~\mathrm{PeV}} \cdot 50$~m) with $E_{\tau}$ being the energy of the
outgoing $\tau$ lepton. The double cascade process could, however, 
manifest itself in the time-stamped waveforms as double pulse. A
search for astrophysical tau neutrino double pulse signature in the waveforms
was carried out with three years of IceCube data. Zero events were found,
consistent with Monte Carlo expectation of $0.5$ $\nu_{\tau}$ double
pulse events in 3 years. For the first time, a differential upper
limit for astrophysical tau neutrinos is set around the PeV energy region
as shown in Fig.~\ref{fig:3yr_tau} \cite{Aartsen:2015dlt}. 
A maximum likelihood analysis of the astrophysical neutrino flux combining
both starting events and through-going track events returned the best fit astrophysical neutrino
flavor ratio to be consistent with equal mixing
\cite{PhysRevLett.114.171102, Aartsen:2015knd}.

 \section{Search for Neutrino Point Sources}
 \label{s3}

\begin{figure*}[t] 
\centering
%% \subfloat{{\includegraphics[width=0.5\textwidth]{7_years_+_MESE_scan_pVal.pdf}}}%
\subfloat{{\includegraphics[width=0.47\textwidth]{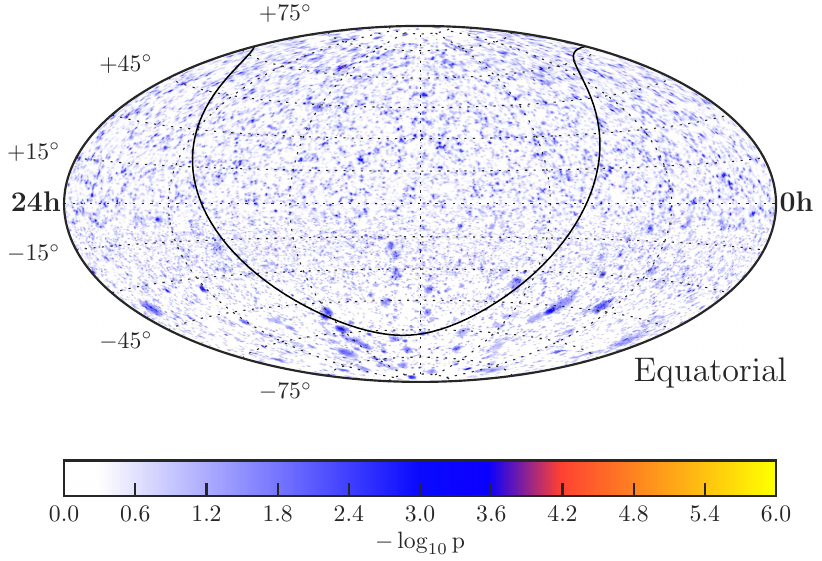}
   }}%
    \qquad
    %%\subfloat{{\includegraphics[width=0.44\textwidth]{sensitivity.pdf}}}%
    \subfloat{{\includegraphics[width=0.47\textwidth]{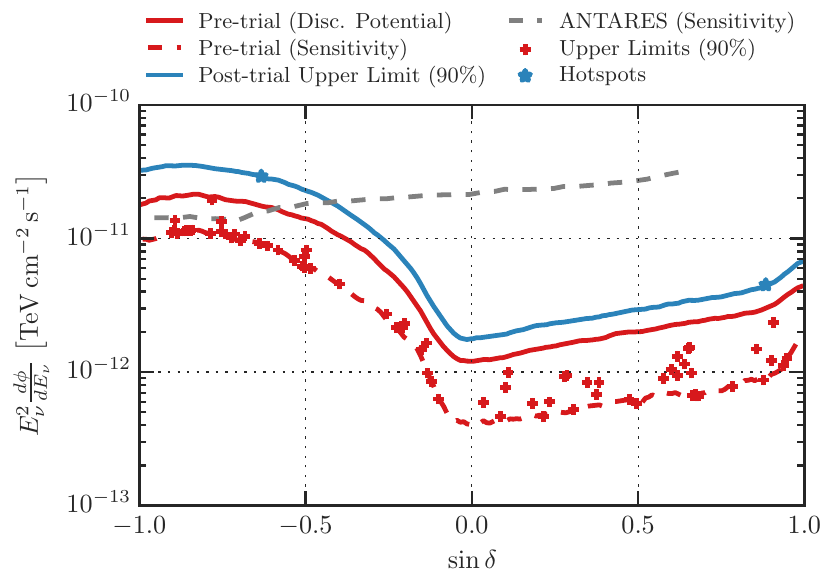} }}%
\caption{Left: results of all-sky time-independent clustering search
  in logarithmic pre-trial p-values. Right: sensitivity (dashed red)
  at 90\% C.L., 5$\sigma$
  discovery potential (solid red) and 90\% C.L. upper limit (blue)
  assuming a $\nu_{\mu} + \bar\nu_{\mu}$ unbroken E$^{-2}$ energy
  spectrum \cite{Aartsen:2016oji}.}%  %%in seven years point source search}.}%
\label{ps_7yr}%
\end{figure*}

Event selections optimized for point source sensitivity and discovery potentials
usually result in looser cuts and larger event samples than those
targeted at diffuse astrophysical flux studies. Point source analyses
in IceCube have better sensitivity in the northern sky using
through-going track-like events which are predominantly atmospheric muon
neutrinos with angular resolution $<1^{\circ}$.
A search for all sky time-independent clustering of
astrophysical neutrinos, using 7 years of data from 2008 to 2015 with
over 700,000 track-like events, did not find any significant steady point like
emission. Both untriggered fine grid full sky scan (suffers from
high trial factor penalty) and correlating to 74 $\gamma$-ray sources (trial factor significantly reduced) were performed. Results are consistent with background and
stringent upper limits on steady astrophysical point source flux are
set on the level of $E^2_{\nu}\frac{d\Phi}{dE_{\nu}} \sim 10^{-12}$
TeV cm$^{-2}$ s$^{-1}$ (Fig.~\ref{ps_7yr}) \cite{Aartsen:2016oji}. Other searches correlate neutrinos to a same category of sources
use a stacked maximum likelihood method to probe the scenario of
distributed weak sources. Such analyses include a recent search for
steady astrophysical neutrino emission from 862 blazars detected in
GeV $\gamma$-rays \cite{Glusenkamp:2015jca}, and
searches for prompt time-correlated neutrinos with 1028 GRBs
\cite{Aartsen:2014aqy, Aartsen:2016qcr}. No significant correlation was found, constraining that less
than 20\% and 1\% of the diffuse flux could have come from blazars and
GRBs, respectively. Complementary to the dedicated
multi-messenger searches, the currently operating realtime alert
system sends high significance alert to partnership observatories
within $O$(min) latency, enabling realtime multi-messenger
follow-ups in the electromagnetic counterparts and
gravitational waves. 

\section{Conclusion and Outlook}
 \label{s4}

The IceCube Neutrino Observatory has detected a diffuse astrophysical
neutrino flux, marking the dawn of neutrino astronomy. Among this flux, a
$\nu_{\mu}$ flux has been observed, while the $\nu_{\tau}$ component
has yet to be identified. No TeV neutrino
point sources have been found at this point. More sophisticated algorithms are
under development to continue the hunt for sources
as more data are collected. A future generation detector in planning, called
IceCube-Gen2, aims to increase the detector volume by a factor of 10 \cite{Aartsen:2014njl}
and make the bottom center sub-array even denser \cite{Aartsen:2014oha}. This will
improve the sensitivity by up to an order of magnitude for both
high and low energy neutrinos. 

The Universe is opaque to the highest energy photons and cosmic rays, but entirely
transparent to neutrinos and gravitational waves. 
Multi-messenger astronomy with electromagnetic signals, cosmic rays,
gravitational waves, and neutrinos will provide unprecedented
opportunities to explore the cosmos.


%%\section{}
%%\label{}

%% References

%%\nocite{*}
\bibliographystyle{elsarticle-num}
%\bibliography{martin}
\bibliography{tau2016}






% \end{thebibliography}

\end{document}